\documentclass[aps,pre,preprint,onecolumn,showpacs,amsmath,amssymb]{revtex4-1}
\usepackage[T1]{fontenc}
\usepackage{amsmath}
\usepackage{graphicx}
\usepackage{bm}
\usepackage{color}
\usepackage{lineno}
\usepackage{soul}
\usepackage{textgreek}
\usepackage[utf8x]{inputenc}
\usepackage{xcolor}

\begin{document}

\title{Microdomains and Stress Distributions in Bacterial Monolayers on Curved Interfaces}

\author{Blake Langeslay$^1$}
\author{Gabriel Juarez$^2$}
\thanks{Email address.}\email{gjuarez@illinois.edu}

\affiliation{$^1$Department of Physics, University of Illinois at Urbana-Champaign, Urbana, Illinois 61801, USA}

\affiliation{$^2$Department of Mechanical Science and Engineering, University of Illinois at Urbana-Champaign, Urbana, Illinois 61801, USA}

\date{\today}


\begin{abstract}

Monolayers of growing non-motile rod-shaped bacteria act as active nematic materials composed of hard particles rather than the flexible components of other commonly studied active nematics. 
The organization of these granular monolayers has been studied on flat surfaces but not on curved surfaces, which are known to change the behavior of other active nematics. 
We use molecular dynamics simulations to track alignment and stress in growing monolayers fixed to curved surfaces, and investigate how these vary with changing surface curvature and cell aspect ratio. 
We find that the length scale of alignment (measured by average microdomain size) increases with cell aspect ratio and decreases with curvature. Additionally, we find that alignment controls the distribution of extensile stresses in the monolayer by concentrating stress in low-order regions. 
These results connect active nematic physics to bacterial monolayers and can be applied to model bacteria growing on droplets, such as marine oil-degrading bacteria.

\end{abstract}

\maketitle

\section{Introduction}

The role of mechanical forces in bacterial growth is of increasing interest to both biologists and physicists \cite{persat, you1, duvernoy, boyer, grant, volfson}. Bacteria colonize a wide variety of interfaces (liquid-solid, liquid-air, liquid-liquid) with vastly different properties \cite{marshall, krajnc, conrad, niepa}. To thrive under these diverse conditions, cells must contend with physical forces such as surface tension and hydrodynamic interactions to stably adhere to the interface over many generations. Bacteria growing on a flat interface as a monolayer of cells have been successfully modeled as an active nematic material. This is a widely studied class of active liquid crystal whose components align parallel to one another and have end-to-end symmetry \cite{you1, dellarciprete}. Rod-shaped bacteria cells have the required symmetry and, when in a dense monolayer, align due to steric interactions between cells. Extensile activity can be produced by motility \cite{copenhagen} or by the growth and division of the cells \cite{dellarciprete}. In either case, forces in the monolayer are strongly coupled to alignment, as both motility and growth exert forces along the axis of orientation.

The active nematic model of bacterial monolayers has proven powerful in predicting the internal forces and behavior of real systems such as monolayers of gliding Myxobacteria and chaining biofilms \cite{copenhagen, yaman}. In particular, the behavior of cells near topological defects has been successfully tied to the behavior of other well-studied active nematic systems. Topological defects are singularities in the director field of liquid crystal alignment, points around which there is a net rotation of the director; this rotation is defined as the defect's charge. In active nematic materials, defects are almost always limited to charges of \(\pm1/2\) \cite{giomi}. These singularities drive much of the unique behavior of active nematics. Comet-shaped +1/2 defects move as motile particles that generate complex flows \cite{giomi, thampi}, and material accumulates at positively charged defects and depletes at negatively charged defects, allowing 2D materials to escape into the third dimension by multilayering or buckling \cite{guillamat, saw, endresen, turiv, kawaguchi}. This last effect has been observed to drive transitions from monolayers to multilayered 3D structures \cite{copenhagen, yaman}.

Previous work on monolayers of rod-shaped bacteria has indicated that when the cells are hard rods (rather than the flexible rods observed in species such as Myxobacteria), their alignment behavior changes. In this case, working with bacteria such as \textit{E. coli}, cells have been observed to segregate into microdomains, regions of near-parallel local alignment analogous to grains in crystalline or granular materials \cite{you1}. Rather than having a continuous, gradually changing alignment field, these systems exhibit sharp changes in alignment across the boundaries between microdomains. This represents a fundamentally different type of liquid crystal behavior from that observed in active nematics composed of microtubules or flexible cells. In these systems, microdomains and their boundaries can replace topological defects as a way of mapping alignment \cite{you1}.
	
The alignment behavior of simulated and microtubule-based active nematic materials is well known to change based on the curvature of their substrate. Topological defects respond to curvature, with $+1/2$ ($-1/2$) defects accumulating at regions of positive (negative) Gaussian curvature \cite{alaimo, ellis, nestler}. However, it is currently unknown what effect curvature might have on the alignment of a more granular hard-cell system. To understand how stresses behave in a curved growing monolayer (for example, one growing on a droplet of oil in water), it is crucial to first understand the alignment behavior of its cells.

To investigate this issue, we simulated the growth of rod-shaped cells on spherical surfaces. This enabled an analysis of how alignment responds to curvature and how stresses in turn respond to alignment. We find that both cell aspect ratio and surface curvature play a role in controlling the length scale of alignment, with higher curvature substrates limiting alignment. Additionally, we find that regions of high stress are predicted by the orientational order, a general measure of low local alignment. These results enable predictions of how granular monolayer behavior might vary on differently curved surfaces.

\section{Methods}

\subsection{Experiments of bacterial growth at flat liquid interfaces}

Cell cultures of \emph{A. borkumensis} were grown for 24 hours in ATCC medium 2698 at 30 $^\circ$C in an orbital shaker at 180 RPM. 
Cells were non-motile and rod-shaped with an average length of 2.7 \textmu m and width of 0.7 \textmu m.
To observe bacterial growth at oil-water interfaces, a custom microfluidic device was used. A flat oil-water interface was pinned to a microscope slide by a thin copper TEM grid (18 \textmu m, SPI Supplies, 2010C-XA) with square apertures 205 \textmu m wide. The cell culture was injected above this, allowing cells to adsorb on the interface. Then, a microfluidic chamber was constructed around the grid to house the interface and allow for constant flow of growth media (diluted 10:1 with artificial seawater) at ~2 \textmu L min$^{-1}$. This flow prevented additional cells from settling on the grid during the experiment.

Time-lapse phase contrast microscopy was used to image the growing cell colony 8 hours using a $60\times$ objective (NA $= 0.6$ and a depth of field of 2 \textmu m). 
A single square aperture in the TEM grid was imaged in an experiment, selected for clarity and lack of visible contaminants. 
Images were recorded with a 50 ms exposure time at 2-minute intervals for 24 hours.

\subsection{Simulations of bacterial growth on flat and curved substrates}

Molecular dynamics simulations were conducted to obtain precise quantitative data on the physical characteristics of growing bacterial monolayers on flat and curved surfaces.
Each cell was modeled as a spherocylinder with a diameter \(d_{0}\) and length \(l\) between the endcaps.
To simulate cell growth, the length of the cylinder \(l\) increased linearly with time up to a maximum length \(l_{0}\) while the diameter remained fixed at \(d_{0}\).
The changes in position \(\vec{x}\) and orientation \(\theta\) of each cell were modeled by the overdamped Newton's equations as follows:
\begin{equation} 
\frac{d\vec{x}}{dt}=\frac{1}{l\zeta}\vec{F}
\end{equation}
\begin{equation} 
\frac{d\theta}{dt}=\frac{1}{l^{3}\zeta}\tau
\end{equation}
where \(\zeta\) is the effective viscosity of the interface based on the interaction between the bacterium and its surroundings and \(\vec{F}\) and \(\tau\) are the total force and torque on the cell due to cell interaction forces.
The interactions between cells were modeled as Hertzian forces, with the force on cell \(i\) due to cell \(j\) calculated as follows:
\begin{equation}
\vec{F_{ij}}=Yd_{0}^{1/2}h_{ij}^{3/2}\vec{N_{ij}}
\end{equation}
where \(Y\) is proportional to the Young's modulus of a cell, \(h_{ij}\) is the overlap distance between the two cell bodies, and \(\vec{N_{ij}}\) is the vector normal to cell \(j\) at the point of contact \cite{orozco-fuentes, you1, hertz}.

Cell growth was modeled using a time-independent rate \(g_{0}\). 
To prevent the synchronized division of cells, each cell was assigned random a value between \(g_{0}/2\) and \(3g_{0}/2\) for the growth rate.
Whenever a cell length $l$ exceeded the maximum length \(l_{0}\), cell division would occur where the cell would split into two identical cells, each with a length \((l_{0}-d_{0})/2\).
The new cell would be initialized with the same orientation $\theta$ as the original cell, however, the new cell would be assigned a different randomized growth rate. 

Simulations of cell growth on flat substrates were initialized with a single parent cell, 
whereas simulations of growth on spherical substrates were initialized with two parent cells, one located at the north pole and the other at the south pole of the sphere. 
This resulted in two hemispherical colonies growing until contact was made at the equator, after which, the distribution of cells rapidly became homogeneous across the entire surface. 
Data was collected on fully-covered spheres once they had reached a packing fraction of \(\phi=1.05\), where \(\phi\) was defined as the area fraction of the surface covered by cells.
In all simulations, the center of volume of cells was constrained to be attached to 
both flat and spherical substrates and the cell orientation was constrained parallel to the surface (or to the tangent plane at the point of contact, for spherical substrates). 
Therefore, no out-of-plane motion was allowed.

Simulation model parameter values were chosen to be representative of a generic gram-negative, rod-shaped bacterium, including those for the \textit{A. borkumensis} cells used in experiments at flat liquid interfaces \cite{you1}.
Therefore, the values were set to the following: cell diameter \(d_{0}\) to 0.7 \textmu m,  Young's modulus \(Y\) to 4 MPa, drag per length \(\zeta\) to 200 Pa h, and  growth rate \(g_{0}\) to 2 \textmu m h$^{-1}$. A simulation time step of $5\times10^{-6}$ hours was used.
To study the effect of different cell aspect ratios, the maximum growth length allowed was 
varied between $2 < l_{0} < 5$ \textmu m.
Cell elongation was parametrized with the aspect ratio $\alpha$, defined here as \(\alpha=(l_{0}+d_{0})/d_{0}\), therefore, ranging from $3.9 < \alpha < 8.1$. 
To study the effect of varying substrate curvature $\kappa$, spherical substrates with radius $R = 10, 12, 15, 20, \ \text{and} \ 30$ \textmu m were used.
Here, substrate curvature is defined as the Gaussian curvature, or \(\kappa = R^{-2}\), for a spherical surface.

\section{Results}

\begin{figure*}
\centering
	\includegraphics[width=\linewidth]{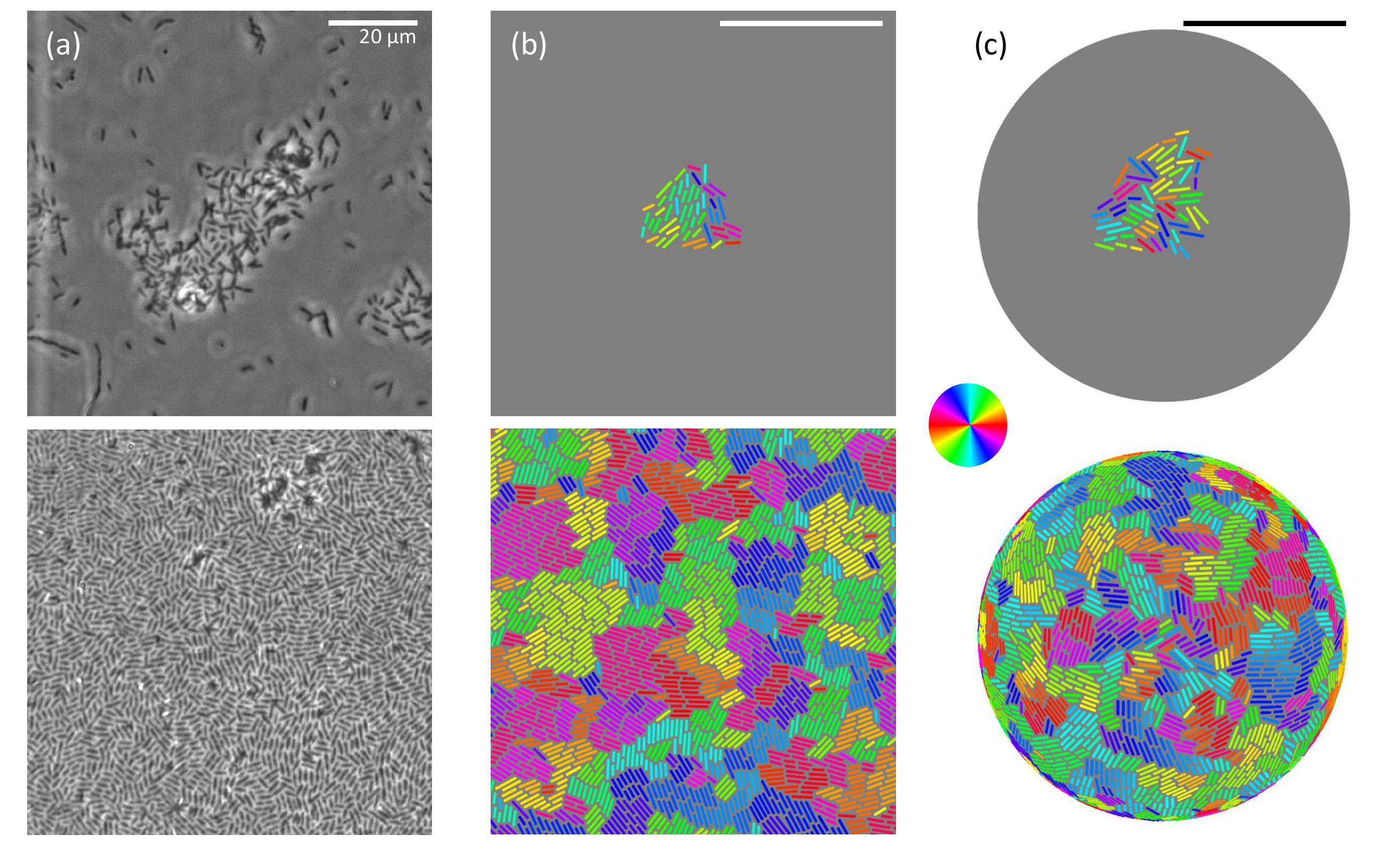}
\caption{
Formation of a bacterial monolayer due to growth at flat and curved substrates. 
(a) Micrographs of \emph{A. borkumensis} cells growing on a flat liquid interface at (top) early times and (bottom) up to full surface coverage.
(b, c) Simulations of cell growth on (b) flat and (c) spherical substrates with $R=20$ \textmu m at (top) early times and (bottom) up to full surface coverage. 
Colors correspond to the cell orientation angle.
All scale bars shown represent 20 \textmu m.
See Supplementary videos SV1, SV2, and SV3.}
	\label{fig:figone}
\end{figure*}

In both experiments and simulations, bacteria grow and divide at the surface, forming a monolayer that eventually covers the entire available surface area, shown in Fig.~\ref{fig:figone}.
At early times, single cells grow and divide to form colonies, shown in Fig. \ref{fig:figone} (top row). 
At later times, as the cells continue to grow and divide, their contact forces and torques cause them to align with their neighbors to form a liquid crystal with nematic symmetry, shown in Fig \ref{fig:figone} (bottom row). 
As cells grow, the area covered by the colony increases exponentially with time until the available surface area is fully covered.
These behaviors are consistently observed in experiments of growth on flat interfaces and in simulations of growth on flat and spherical substrates.

To establish correspondence between experiments and simulations, the distribution of topological defects in cell monolayers on flat interfaces and substrates, respectively, are compared. 
First, to identify topological defects, a director field was established. 
In experiments, the director field was determined using a custom image-processing algorithm based on the brightness gradient of the phase contrast images.
In simulations, the director field was generated directly from the position and orientation of each cell. 
Then, to locate topological defects within the director field, each point on the grid (or each cell in simulations) was tested for a net rotation of the surrounding director field (with net rotations of $\pm \pi/2$ corresponding to defect charges of $\pm 1/2$) \cite{decamp}.
Nearby points with similar net rotations were then separated into clusters, with the centroid of each cluster corresponding to a defect of the associated charge.

\begin{figure*}
\centering
	\includegraphics[width=\linewidth]{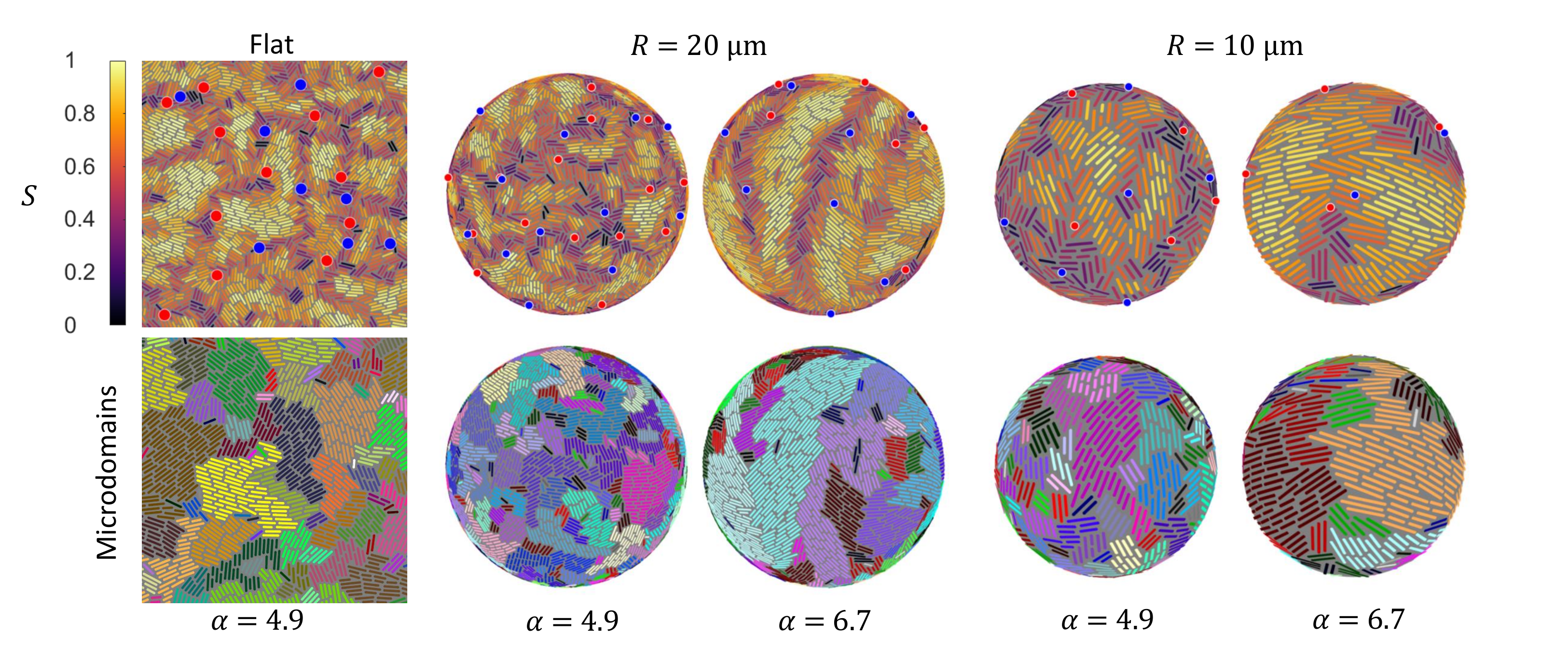}
\caption{
Orientational order, topological defects, and microdomains in simulations of monolayers with varying cell aspect ratio \(\alpha\) and substrate radii \(R\). 
(Top row) Orientational order \textit{S} and topological defects with charge $\pm 1/2$ shown in red/blue, respectively. 
Topological defects tend to occur in regions of low orientational order for all aspect ratios and substrate radii.
(Bottom row) Microdomain representation of the cell monolayers shown in the top row.
Microdomain boundaries tend to coincide with regions of low orientational order for all aspect ratios and substrate radii. 
For visualization purposes, spherical substrates ($R=10$ \textmu m and $R=20$ \textmu m) are not shown to scale while the field of view for the flat simulations (leftmost column) is $50 \times 50$ \textmu m.
}
	\label{fig:figtwo}
\end{figure*}

The distribution of topological defects in experiments at flat interfaces corresponds to simulations of growing cells with the same dimensions as \textit{A. borkumensis} on flat substrates, shown in Fig.~\ref{fig:figtwo} (top row, flat) and Supplementary Fig. S1. 
This is determined using the mean defect separation, defined as the average over the three nearest-neighbor distances for all defects.
In experiments, the mean defect separation is $8.9 \pm 3.1$ \textmu m, while 
in simulations, the mean defect separation is $8.17 \pm 2.9$ \textmu m.

Based on the good qualitative and quantitative agreement between experiments and simulations, the effect of cell aspect ratio $\alpha$ and substrate curvature $\kappa$ on orientational order, the degree of alignment, and stress within the cell monolayer was investigated using simulations.
First, the orientational order \textit{S} was evaluated to measure the degree of local alignment in the monolayer.
This was calculated at each individual cell \textit{i} using the following equation:
\begin{equation}
    S_{i}=\sum_{j} \frac{1}{2}(3\cos^2(\theta_{i}-\theta_{j})-1)
\end{equation}
where \(\theta_{j}\) is the orientation of each cell within a search radius of cell \(i\). For the purpose of this analysis, the search radius was set equal to the division length \(l_{0}\) of the cells. 
Here, $S \rightarrow 1$ represents ordered regions while $S \rightarrow 0$ represents disordered regions.

Simulations reveal regions of high order separated by regions of low order, shown in Fig. \ref{fig:figtwo} (top row). 
These regions of high cell alignment emerge for all combinations of substrate curvature and cell aspect ratio, including flat substrates. 
For a given curvature, increasing the cell aspect ratio from $\alpha = 4.9$ to $\alpha = 6.7$ increases the size of these regions, shown in Fig. \ref{fig:figtwo} ($R=20$ \textmu m). 
Additionally, topological defects tend to coincide with areas of low order, shown in Fig.~\ref{fig:figtwo} (top row). 
This follows from the definition of defects, as the net rotation of the director field around the defect requires imperfect alignment. 
Therefore, increasing the cell aspect ratio also increases the distance between $\pm$ 1/2 defect separation, as shown in Supplementary Fig. S2.

Next, the mean microdomain area \(\langle A \rangle\) was calculated to characterize the degree of cell alignment in the monolayer.
Boundaries between microdomains are one-dimensional discontinuities in the alignment field rather than point defects, separating regions of near-parallel alignment. 
Cells were sorted into microdomains using the following two criteria: (i) cells were in contact with one another and (ii) their orientation differed by less than 0.2 radians.

Microdomains, represented as differently colored regions, correspond to regions of high-aligned cells, shown in Fig. \ref{fig:figtwo} (bottom row).
Borders between large microdomains correspond to regions of low order, reflecting the discontinuity in alignment between microdomains.
Similarly, for a given curvature, increasing the cell aspect ratio increases the size of a single microdomain, shown in Fig. \ref{fig:figtwo} ($R=20$ \textmu m).
The area of a microdomain is given as $A=\phi\sum{A_{i}}$, where $A_{i}$ are the areas of each cell component within the microdomain. 
For all substrate curvatures and aspect ratios investigated here, the distribution of microdomain areas within the monolayer is described by $P(A) \propto$ exp($A/ \langle A \rangle$), where $\langle A \rangle$ is the mean microdomain area for that monolayer, shown in Supplementary Fig. S3. \cite{you1}. 
The mean microdomain area for a system is used to characterize the area scale of its cell alignment.

\begin{figure*}
\centering
	\includegraphics[width=\linewidth]{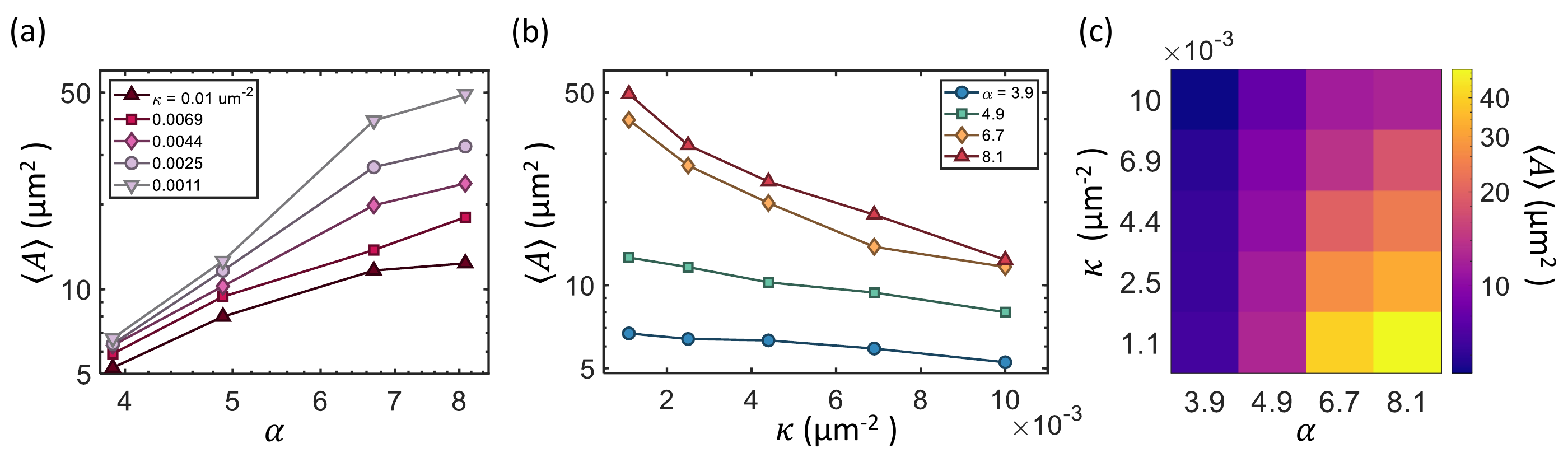}
\caption{
The effect of cell aspect ratio and substrate curvature on microdomain area. 
(a) Log-log plot of mean microdomain area \(\langle A \rangle\) for different cell aspect ratios \(\alpha\) on five different substrate curvatures. 
Microdomain area increases with the aspect ratio for all curvatures. 
(b) Semilog plot of mean microdomain area \(\langle A \rangle\) for different substrate curvatures \(\kappa\) and four different cell aspect ratios. 
Microdomain area decreases with substrate curvature for all aspect ratios. 
(c) Summary of the mean microdomain area \(\langle A \rangle\) across the simulated parameter space of aspect ratio and substrate curvature.
}
	\label{fig:figthree}
\end{figure*}

The mean microdomain area increased with increasing cell aspect ratio, shown in Fig. \ref{fig:figthree}(a). 
For the lowest curvature ($\kappa=0.001$, $R=30$ \textmu m) the effect of cell aspect ratio was the most dramatic, with an increase in the mean domain area of seven-fold. 
For the highest curvature ($\kappa=0.01$, $R=10$ \textmu m), however, the effect of cell aspect ratio was less prominent, producing an increase in the domain area of a factor of two. 
In fact, the dependence of mean domain area on aspect ratio could be described with a power law relation in the range of aspect ratios investigated, with scaling exponents increasing from 1.2 for $\kappa=0.01$ up to 2.9 for $\kappa=0.001$. 
The increase in alignment at higher \(\alpha\) is consistent with previous work on bacterial monolayers, which has shown that more elongated (higher aspect ratio) cells produce stronger alignment in flat monolayers \cite{you1}.

The mean microdomain area decreased with increasing substrate curvature, shown in Fig. \ref{fig:figthree}(b).
For low aspect ratios ($\alpha \leq 5$), an inverse exponential form with $\langle A \rangle \propto$ exp($-\kappa/\kappa_{0}$) accurately describes the relation between area and curvature.
Specifically, for $\alpha$ of $3.9$ and $4.9$, the value of $\kappa_0$ was equal to $0.039$ \textmu m$^{-2}$ and $0.020$ \textmu m$^{-2}$, respectively. 
For high aspect ratios ($\alpha > 5$), the inverse exponential does not accurately describe the relation between area and curvature, demonstrating a qualitative change in the system's alignment behavior.
In all cases, however, more curved substrates produced consistently lower microdomain areas.

Measurements of mean domain area across a range of aspect ratios and curvatures were combined to show the system's response over the $\alpha - \kappa$ parameter space, shown in Fig. \ref{fig:figthree}(c). 
The combined effect of cell aspect ratio and substrate curvature on microdomain area is evident. 
The largest domain areas are observed at high cell aspect ratios and low substrate curvatures.
The smallest domain areas, however, are observed at low cell aspect ratios and high substrate curvatures.

Lastly, the parallel component of the Virial stress \(\sigma_{\parallel}\) on each cell was measured to determine the force distributions in the monolayer. 
The Virial stress \(\boldsymbol{\sigma}_{i}\) on a cell is given as follows:
\begin{equation}
    \boldsymbol{\sigma}_{i}=\frac{\phi}{a_{i}}\sum_{j} \boldsymbol{r}_{ij}\boldsymbol{F}_{ij}
\end{equation}
where \(a_{i}\) is the area of cell \(i\), \(\boldsymbol{r}_{ij}\) is the vector from the center of cell \(i\) to the point of contact with cell \(j\), and \(\boldsymbol{F}_{ij}\) is the force from cell \(j\) on cell \(i\). 
When \(\boldsymbol{\sigma}\) is calculated in the basis of vectors parallel and perpendicular to the cell's orientation, it can be decomposed into parallel ($\sigma_{\parallel}$), perpendicular ($\sigma_{\perp}$), and shear (\(\tau\)) components:
\begin{equation*}
    \boldsymbol{\sigma}=\begin{bmatrix}
    \sigma_{\parallel} & \tau/2\\
    \tau/2 & \sigma_{\perp}
    \end{bmatrix}
\end{equation*}
The parallel stress \(\sigma_{\parallel}\) corresponds to the force in the direction of the cells' growth. 
Because of the extensile nature of the system, \(\sigma_{\parallel}\) is always negative (corresponding to a compressive force) \cite{you1}.

\begin{figure*}
\centering
	\includegraphics[width=\linewidth]{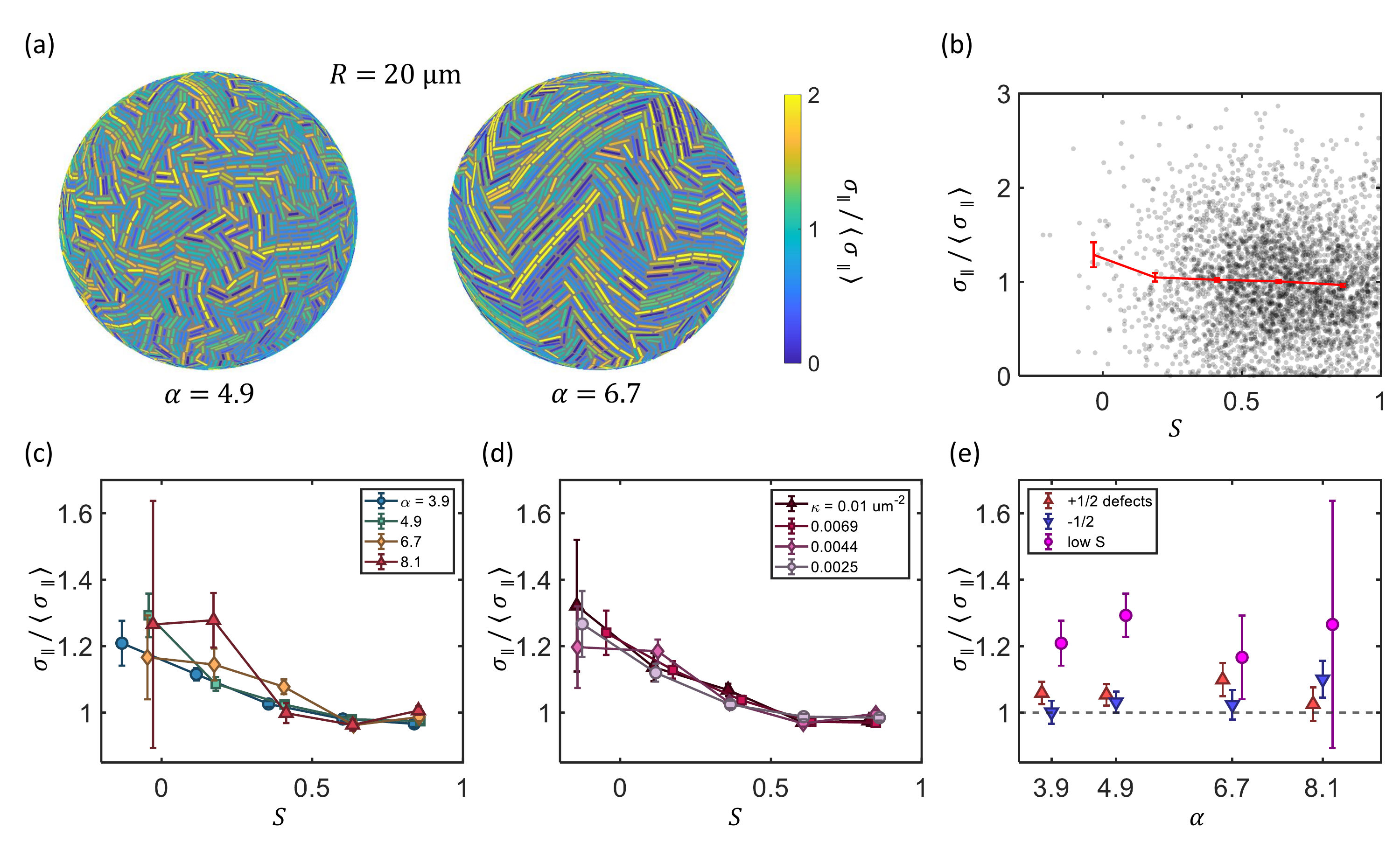}
\caption{
Parallel component of the Virial stress $\sigma_{\parallel}$ in growing cell monolayers on curved surfaces. 
(a) Visualization of the normalized parallel stress in cell monolayers on a curved surface with $R=20$ \textmu m and for cell aspect ratios of $\alpha=4.9$ and $6.7$. 
(b) Scatter plot of the normalized parallel stress of individual cells and their local orientational order for the case of $R=20$ \textmu m and $\alpha=4.9$.
The trendline shows the data binned by orientation order and the errorbars represent the standard error.
(c) Relation between the normalized parallel stress magnitude and the orientational order for differing aspect ratios. 
Average parallel stress magnitude increases by up to 30\% in the lowest-order bin ($S<0.1$). 
(d) Relation between the parallel stress magnitude and the orientational order for differing surface curvatures.
(e) Deviation from mean parallel stress at topological defects ($+1/2$ in red, $-1/2$ in blue) and in low-order regions ($S<0.1$ in magenta). 
Low-order regions have a greater deviation from the mean stress than topological defects for all curvatures and aspect ratios.
}
	\label{fig:figfour}
\end{figure*}

The parallel component of the Virial stress, normalized by the mean stress, for each cell in the monolayer on a curved substrate with $R=20$ \textmu m is visualized and shown in Fig. \ref{fig:figfour}(a). 
At the cell level, the normalized stress varied over a wide range of values.
For example, for $\alpha=4.9$ and $R=20$, the mean stress was $\langle \sigma_{\parallel} \rangle =-0.043$ N m and the normalized stress varied from a minimum of zero up to a maximum of $\sim 3.8 \langle \sigma_{\parallel} \rangle$.
For an aspect ratio of $\alpha=6.7$ and the same surface curvature, the mean stress was $\langle \sigma_{\parallel} \rangle =-0.035$ N m and the normalized stress varied from a minimum of zero up to a maximum of $\sim 4.4 \langle \sigma_{\parallel} \rangle$. 
Stress in growing monolayers with varying values of $\alpha$ and $\kappa$ can be seen in Supplementary videos SV4-SV7.

A scatter plot of the normalized parallel component of Virial stress and the orientational order for a representative simulation ($R=20$, $\alpha=4.9$) is shown in Fig. \ref{fig:figfour}(b), grey points.
It is evident that, at the individual cell level, stress values can vary over a wide range with respect to the average value of stress. 
While no trend is immediately visible in the scattered data, binning the data by the orientational order $S$ reveals a relationship between the normalized stress $\sigma_{\parallel}/\langle \sigma_{\parallel} \rangle$ and orientational order $S$.
First, five evenly spaced bins were determined by identifying the minimum $S_{min}$ and maximum $S_{max}$ values of the orientational order for each simulation. 
Then, the mean orientational order and mean parallel stress were computed for each bin. 
The result is shown in Fig. \ref{fig:figfour}(b), red line. 
For this example ($R=20$, $\alpha=4.9$), the binned data shows that cells in regions of low-order experience a higher magnitude of parallel stress compared to cells in high-order regions.

The trend of higher-magnitude stress at lower-order regions is a robust observation for all simulations with varying cell aspect ratio and substrate curvature. 
Plots of normalized parallel stress binned by order and averaged across all simulations with the same aspect ratio or surface curvature are shown in Fig. \ref{fig:figfour}(c) and \ref{fig:figfour}(d), respectively. 
For all cases, the average normalized stress decreases with increasing orientational order.  
Specifically, in regions of high orientational order ($S>0.5$), the local average of normalized stress approaches unity, or $\sigma_{\parallel} \approx \langle \sigma_{\parallel} \rangle$.
In regions of low orientational order, however, ($S<0.1$), the local average of normalized stress is 15\% to 35\% larger than the global average stress, or $\sigma_{\parallel} \geq 1.15 \langle \sigma_{\parallel} \rangle$.
The reason for the large standard error for $\alpha = 8.1$ and $S<0.1$ is that the high aspect ratio cells are more aligned on average, resulting in a sparse number of cells in low-order regions to analyze.

The stress in other active nematics such as those composed of epithelial cells is higher near $+1/2$ topological defects, which themselves are low-order regions by definition \cite{guillamat}. 
To investigate whether this effect was responsible for the correlation of high stress and low order in bacterial monolayers, stress near topological defects was compared to the previously calculated stress in low-order regions ($S<0.1$).
The mean value of the parallel component of the Virial stress ($\sigma_{\parallel}$) was computed for all cells within close proximity ($r=l_{d}/2$) of a $\pm 1/2$ topological defect within the monolayer. Then, this value was compared to the average normalized parallel stress of the bin with the lowest orientational order ($S<0.1$) and is shown in Fig. \ref{fig:figfour}(e).
For all cases, the average deviation from the mean stress near $\pm 1/2$ defects is less than 10\%, or $\sigma_{\parallel} \leq 1.1 \langle \sigma_{\parallel} \rangle$.
In contrast, the average deviation from the mean stress in regions of low orientational order ($S<0.1$) is always greater than 15\%, or $\sigma_{\parallel} \geq 1.15 \langle \sigma_{\parallel} \rangle$, for all cell aspect ratios and substrate curvatures.
This shows that the observed high stress-low order correlation in this system is not caused by stress concentration near topological defects, and is instead a distinct effect.

\section{Discussion}

Previous work on monolayers of growing hard-rod cells has shown that, in the case of a flat substrate, microdomain size increases with increasing cell aspect ratio \cite{you1}. 
Our work is consistent with this result, and further confirms that the trend holds for monolayers growing on curved substrates as well. 
Additionally, while previous work demonstrated the relation between microdomain size and aspect ratio for an unconfined growing colony \cite{you1}, our results show that the same relation is true for a growing colony confined to a finite substrate area (the surface of a sphere). 
Together, these show that microdomain formation and its dependence on cell aspect ratio are robust collective behaviors in growing hard-rod monolayers under a variety of conditions.

We also find a new dependence of microdomain area on the curvature of the surface. 
Regardless of cell aspect ratio, higher curvature substrates resulted in smaller microdomains. 
This decrease in alignment is attributable to the surface’s curvature restricting cells from lying parallel to each other. 
Microdomains can only maintain close cell alignment over a small enough area to be approximated as locally flat, and a microdomain spanning a much larger area on a curved surface will eventually be geometrically required to fracture. 
As curvature increases, the maximum area that can be treated as locally flat decreases, and accordingly the domain area decreases as well.

In a continuous (non-granular) active nematic composed of flexible epithelial cells, extensile stress is highest at positively charged topological defects \cite{guillamat}.
This causes the cell layer to deform at these points, producing “mounds” localized near the defects\cite{guillamat}. 
Similar deformations are seen in other continuous active nematics at deformable 2D interfaces, such as microtubules on the surface of a vesicle or actin fibers in the morphogenesis of multicellular organisms \cite{keber, maroudas-sacks}, and this behavior has been further confirmed and studied in numerous simulations and theoretical works \cite{metselaar, ruske, santiago, vafa, hoffmann, alert}. 
In a granular active nematic, our simulations show that extensile stress concentrates at all low-order regions rather than at the locations of point $+1/2$ topological defects. 
This can be expected to result in different forms of deformation under growth stress.

Our results are relevant for systems of bacteria growing at liquid-liquid interfaces, such as at the oil-water interface of a droplet. 
For example, previous work has shown that cell growth confined to the surface of a droplet produces tube-like protrusions similar to those produced by continuous active nematics \cite{hickl1, prasad}. 
However, a complete theoretical description of droplet deformation by cell growth is still lacking.
Our results suggest that these protrusions should nucleate at low-order sites such as boundaries between microdomains, rather than nucleating exclusively at $+1/2$ defects. 
By extension, this allows us to predict how interfacial curvature influences deformations due to the underlying microstructure. 
For example, higher curvature (smaller) droplets will produce more closely spaced protrusions since the characteristic size of their microdomains decreases.

Our results also emphasize the importance of constituent particle properties on collective behavior. 
While a hard-rod monolayer exhibits many of the same properties as a nematic composed of microtubules or other flexible components \cite{dellarciprete}, its internal forces are not well predicted by topological defects.
Furthermore, a growing self-organized monolayer produces microdomains that are distinct from the continuous alignment fields of other active nematic systems. 
The partially granular nature of a bacterial monolayer is clearly of great importance to understanding its collective behavior.

In conclusion, we have shown that stress distributions in a hard-rod bacterial monolayer vary predictably based on cell aspect ratio and substrate curvature. 
Specifically, our experimentally validated simulations show that stress in a hard-rod monolayer concentrates at low-order regions, which occur at the boundaries of microdomains. 
The length scale of these microdomains increases with cell aspect ratio and decreases with substrate curvature. 
These results demonstrate that while a bacterial monolayer can be effectively modeled as a continuum active nematic, in some cases when its cells act as hard rods, the alignment and stress distributions behave in distinctly different ways.

\section{Conflicts of Interest}

There are no conflicts of interest to declare.

\section{Acknowledgements}

This work used the eXtreme Science and Engineering Discovery Environment (XSEDE), which is supported by National Science Foundation grant number \#ACI-1548562. 
In particular, we used the Pittsburgh Supercomputing Center's Bridges-2 resources under allocation ID PHY210132. 
We thank Vincent Hickl for assistance with experiments on growing bacterial monolayers at liquid interfaces.

\bibliography{microdomains}

\end{document}